\documentclass[11pt]{article}
\usepackage[T1]{fontenc}          
\usepackage{amsmath}
\usepackage{amssymb}
\usepackage{geometry}
\usepackage{graphicx}             
\usepackage{caption}              
\usepackage{subcaption}           
\usepackage{hyperref}             
\usepackage{booktabs}   
\usepackage{array}      

\title{Higgs Boson Production in Association with a Single Top Quark as a Probe of the Top Yukawa Coupling}
\author{Tetiana Obikhod and Ievgenii Petrenko \\ \textit{Institute for Nuclear Research NAS of Ukraine, Kyiv, Ukraine}}
\date{}

\begin{document}

\maketitle

\begin{abstract}
\noindent This paper provides a detailed analysis of the associated production of the Higgs boson with a single top quark ($tH$) in proton-proton collisions at $\sqrt{s} = 13~\mathrm{TeV}$ and $14~\mathrm{TeV}$. Based on the ATLAS search, we have employed innovative modeling approaches to improve the sensitivity to new physics and improve Standard Model  constraints. The major goals are to optimize the data selection, statistical error estimation, determination of physical limits on the top quark Yukawa coupling ($\kappa_t$), and future experimental projections for HL-LHC. Simulations with \textsc{MadGraph5\_aMC@NLO} at LO+MLM give cross-sections of $\sigma_{tHq}$ and $\sigma_{tWh}$ in SM, scaled by K-factors to simulate NLO accuracy. For inverted $\kappa_t = -1$ (ITC), positive enhancements are seen, consistent with constructive interference. Kinematic distributions ($H_T$, $p_T(h)$, $\eta[j]$, $\Delta R[t,h]$) are checked against ATLAS expectations, confirming the optimized approach to maximize signal extraction and suppress systematics.
\end{abstract}
\section{Introduction} The observation of the Higgs boson in 2012 by the ATLAS and CMS experiments represented a major milestone in particle physics, confirming the Standard Model ($\mathrm{SM}$) mechanism of mass generation. The couplings of the Higgs boson to other particles, proportional to their masses, are essential for testing the $\mathrm{SM}$. The production process with a single top quark ($tH$) provides a unique window into the top-Higgs Yukawa coupling ($y_t$), which is sensitive to new physics effects from destructive interference in the $\mathrm{SM}$ diagrams.

\begin{figure}[htbp]
\centering
\includegraphics[width=0.8\textwidth]{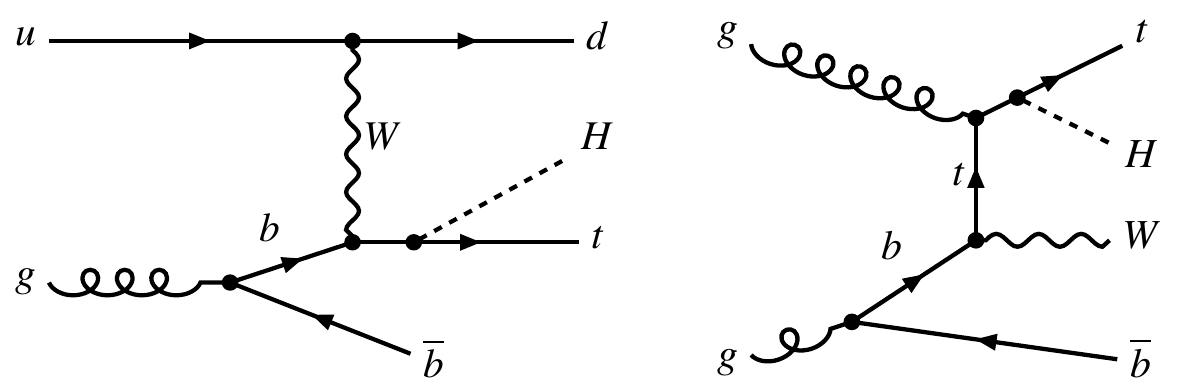} 
\caption{Representative dominant leading-order Feynman diagrams in the 4FS: (left) $tHq$ production via $t$-channel-like exchange and (right) $tWH$ production via gluon $b$ interaction with time-like virtual top quark mediator and Higgs radiation, ~\cite{atlas_tH_2025}}
\label{fig:1}
\end{figure}

The production cross-section of the Higgs boson associated with a single top quark
is sensitive to both the magnitude and the sign of $y_t$. The magnitude and the sign of the
top Yukawa coupling for $t\bar{t}H$ and single top production are sensitive to different kinds of
new physics~\cite{tait2000single,barger2010single}.
Similar sensitivity is obtained from the combination of gluon-gluon fusion (ggF),
vector-boson fusion (VBF), and associated production with vector bosons (VH)~\cite{atlas2022map,cms2022portrait}. The origin of the $b$ quark in the hard process depends on the flavour scheme (FS) employed in the perturbative calculation. In the five-flavour scheme (5FS), the $b$ quark is treated as a massless parton from the proton sea (via DGLAP-evolved bottom PDFs). In contrast, in the four-flavour scheme (4FS), $b$ quarks are absent from the initial-state parton distributions of the proton and are instead produced perturbatively through gluon splitting into a $b\bar{b}$ pair~\cite{maltoni2012binitiated}. In the $tWH$ process, a gluon from one of the protons interacts with a $b$-quark from the other proton or from gluon splitting, mediated by the exchange of a time-like virtual top quark $t$. Two examples of the dominant leading-order (LO) Feynman diagrams for the $tHq$ and $tWH$ processes, both within the 4FS, are given in Figure~\ref{fig:1}.

Higgs boson production accompanied by a single top quark takes place via the electroweak (EW) interaction and includes three different production modes: the $t$-channel ($tHq$), the $\bar{t}t q$ mode (associated production with a top-quark pair and an additional light quark), and the associated $tWH$ mode. Single top quark production takes place via the EW interaction in proton-proton ($pp$) collisions within the Large Hadron Collider (LHC)~\cite{LHC2008}. In this paper, the $\bar{t}t q$ and $tWH$ modes will be collectively referred to as the $tH$ mode, which is used for the measurements, whereas the $s$-channel mode is not taken into consideration due to its very low production cross-section, which amounts to approximately 3\% of the total $tH$ cross-section at LHC energies.

The production cross-sections for the different modes depend on the collision energy and the centre-of-mass energy $\sqrt{s}$. The production cross-section for the $\bar{t}t q$ mode at $\sqrt{s} = 13$ TeV in the Standard Model is approximately 2.4--2.9~fb (depending on the exact calculation and PDF set), while the $tWH$ mode has a significantly smaller cross-section of about 0.07--0.1~fb. The $t$-channel $tHq$ mode is the dominant contribution to single-top + Higgs production, with a cross-section of order 70--90~fb at $\sqrt{s} = 13$ TeV~\cite{atlas_tH_2025, cms_tH}.

These values are sensitive to the magnitude and sign of the top-quark Yukawa coupling $y_t$ (or the effective coupling modifier $\kappa_t$), particularly in the $tWH$ and $t$-channel modes where interference effects between different diagrams can lead to constructive or destructive contributions depending on the relative sign between the top and $W$-boson couplings to the Higgs boson. In the case of a negative (inverted) sign of $y_t$ relative to the vector couplings, the $tWH$ cross-section can be enhanced by up to a factor of $\sim$5--10 compared to the Standard Model expectation due to constructive interference.

The following table summarises approximate Standard Model cross-sections for single-top + Higgs associated production at $\sqrt{s} = 13$ TeV (values taken from recent ATLAS and theoretical predictions):

\begin{table}[htbp]
\centering
\caption{Approximate Standard Model production cross-sections for single top + Higgs associated modes at $\sqrt{s} = 13$ TeV.}
\begin{tabular}{lcc}
\hline
Mode          & Process description                  & Cross-section (fb) \\
\hline
$t$-channel   & $tHq$ (dominant)                     & 70--90            \\
$\bar{t}t q$  & $pp \to \bar{t}t q H$                & 2.4--2.9          \\
$tWH$         & $pp \to t W^\pm H$                   & 0.07--0.1         \\
$s$-channel   & $t H q'$ (neglected)                 & $\sim$0.01--0.03  \\
\hline
\end{tabular}
\label{tab:xsec}
\end{table}

The production cross-section for the $\bar{t}t q$ mode at higher centre-of-mass energies (e.g., $\sqrt{s} = 14$ TeV or future HL-LHC projections) is expected to increase roughly logarithmically with energy due to the gluon-dominated initial state and the scaling of the parton luminosities.

Therefore, this research also aims to contribute significantly to the understanding of Higgs sector interactions with quarks, which remains of fundamental importance for confirming (or potentially challenging) the existing paradigms in particle physics. Early searches by the ATLAS experiment, based on Run~1 data ($\sqrt{s}=7$ and $8$~TeV), provided upper limits on the single top + Higgs production cross-section. Subsequent Run~2 analyses at $\sqrt{s}=13$~TeV, utilizing the full dataset of approximately 140~fb$^{-1}$, achieved first evidence and later observation of $tH$ production in the period 2018--2020. The measured signal strengths were found to be consistent with Standard Model predictions within experimental uncertainties~\cite{atlas_tH_2025}. Projections for the High-Luminosity LHC (HL-LHC), with an expected integrated luminosity of 3000--4000~fb$^{-1}$ at $\sqrt{s}=14$~TeV, anticipate precise measurements of the top Yukawa coupling modifier $\kappa_t$ (or directly $y_t$) with a relative precision of approximately 5--10\%, significantly improving our knowledge of the Higgs--top quark interaction and potentially revealing deviations from the Standard Model. This introduction sets the stage for the subsequent discussions on experimental procedures, analysis strategies, and phenomenological implications for Higgs physics and beyond-the-Standard-Model scenarios.

\section{Analysis Strategy}

The production of a Higgs boson in association with a single top quark ($tH$ production) provides a powerful probe of fundamental aspects of the SM of particle physics, in particular the top-quark Yukawa coupling $y_t$ (or the effective coupling modifier $\kappa_t = y_t / y_t^{\rm SM}$). Any significant deviation from the SM prediction could indicate the presence of new physics beyond the Standard Model.

In the SM, Higgs production in association with a single top quark proceeds mainly through the $t$-channel ($tHq$) and the associated $tWH$ mode. At leading order, these processes exhibit destructive interference between diagrams involving the Higgs-top Yukawa coupling and those mediated by $W$-boson exchange. This interference suppresses the cross-sections, resulting in relatively small values at next-to-leading order (NLO) in QCD for $\sqrt{s} = 13$ TeV, Table 2.

\begin{table}[htbp]
\centering
\caption{Approximate NLO cross-sections for single top + Higgs production at $\sqrt{s} = 13$ TeV in the SM and in the inverted top coupling (ITC, $\kappa_t = -1$) scenario.}
\begin{tabular}{lcc}
\hline
Process       & SM ($\kappa_t = +1$) & ITC ($\kappa_t = -1$) \\
\hline
$t$-channel ($tHq$) & $\sim 74.3$ fb & $\sim 740$ fb \\
$tWH$               & $\sim 15.2$ fb & $\sim 150$ fb \\
Total $tH$          & $\sim 89.5$ fb & $\sim 890$ fb \\
\hline
\end{tabular}
\label{tab:xsec_th}
\end{table}

In the Inverted Top Coupling (ITC) scenario, where the relative sign of the top Yukawa coupling is flipped ($\kappa_t = -1$), the destructive interference turns constructive. This leads to a dramatic enhancement of the $tH$ production cross-section by a factor of approximately 10, yielding

\[
\sigma_{tH}^{\rm ITC} \approx 890~\text{fb}
\]

at the same centre-of-mass energy. Such a large increase makes $tH$ production particularly sensitive to the sign of $y_t$ and provides one of the most direct ways to constrain (or potentially discover) non-standard Higgs-top interactions. The analysis strategy therefore focuses on maximizing the sensitivity to this interference pattern through:
\begin{itemize}
  \item Optimized event selection in final states with high lepton multiplicity, $b$-tagged jets, and missing transverse energy (for $tWH$ and $t$-channel signatures),
  \item Multivariate techniques (boosted decision trees, neural networks) to discriminate signal from dominant backgrounds (mainly $t\bar{t}$ + jets, single top, $W$/$Z$ + jets),
  \item Combination of different decay channels of the Higgs boson ($H \to b\bar{b}$, $H \to WW^*$, $H \to \tau^+\tau^-$, $H \to \gamma\gamma$, etc.),
  \item Interpretation in terms of effective coupling modifiers $\kappa_t$ and $\kappa_V$ (vector boson coupling), including CP-sensitive observables and differential distributions.
\end{itemize}

These strategies have already enabled ATLAS and CMS to achieve evidence for $tH$ production and to set strong constraints on negative values of the relative top Yukawa coupling.

Recent experimental searches performed by the ATLAS Collaboration, employing the complete Run-2 dataset of 140 fb$^{-1}$ at the LHC ($\sqrt{s}=13$ TeV), revealed an excess of $tH$ production. Under the Standard Model hypothesis, a signal strength of $\mu_{tH} = 8.1 \pm 2.6 \, (\text{stat.}) \pm 2.0 \, (\text{syst.})$ is observed, corresponding to a local significance of 2.8$\sigma$ (with an expected significance of 0.4$\sigma$)~\cite{atlas_tH_2025}. This excess is compatible with the interpretation of an inverted top coupling (ITC) scenario ($\kappa_t = -1$), for which a signal strength of $\mu_{tH} = 1.2 \pm 0.4 \, (\text{stat.}) \pm 0.5 \, (\text{syst.})$ is obtained. The analysis sets observed (expected) upper limits at 95\% confidence level of 13.9 (6.1) times the SM prediction for the SM hypothesis, and 2.4 (1.2) times the ITC prediction for the ITC hypothesis. These results, while showing a weak but intriguing excess in the SM fit, motivate further investigation to reduce modeling uncertainties, validate simulation tools (including parton showering, PDF sets, and higher-order corrections), and extend calculations to higher energy scales (e.g., HL-LHC at $\sqrt{s}=14$ TeV) or alternative BSM scenarios with modified Higgs-top interactions.

By linking our computed data---optimized for the four-flavour scheme (4FS) in the $tHq$ channel and the five-flavour scheme (5FS) in the $tWH$ channel, with dynamic factorization and renormalization scales and ATLAS-inspired event selection cuts---to the published experimental observations, we aim to achieve the following objectives:

\begin{enumerate}
  \item Verify the consistency of our MadGraph5\_aMC@NLO setups against the ATLAS modeling choices, including parton showering (Pythia8/Angantyr), PDF sets, and matching/merging procedures;
  
  \item Estimate the theoretical uncertainties associated with leading-order (LO) + MLM (MadGraph merging) approximations for $tH$ production cross-sections and key kinematic distributions (e.g., $p_T(H)$, $p_T(t)$, angular correlations);
  
  \item Investigate the Inverted Top Coupling (ITC) scenario by inverting the sign of the top Yukawa coupling ($y_t \to -y_t$ or $\kappa_t = -1$), which could potentially account for the observed excess in the ATLAS $tH$ search;
  
  \item Extrapolate the results to future LHC runs (Run~3 and High-Luminosity LHC) and explore implications for selected beyond-the-Standard-Model (BSM) scenarios with modified Higgs-top interactions.
\end{enumerate}

This systematic bridge between state-of-the-art theoretical predictions and experimental observations underpins the central goals of our investigation: to enhance the theoretical accuracy and reliability of $tH$ modeling in current and future searches, and to contribute to a deeper understanding of the intriguing excess reported by ATLAS in the single top + Higgs channel.
Our analysis strategy is based on the following key elements:

\subsection{Kinematics and Topology}

In the context of high-energy proton-proton scattering, the $tHq$ process proceeds via the $t$-channel mechanism of single top quark production. The hard scattering is initiated by an initial light quark (predominantly $u$, $d$, $c$, or $s$ flavors from the proton PDFs) interacting with a $b$ quark, mediated by the exchange of a virtual $W$ boson. This process is accompanied by a forward spectator jet, which is typically emitted at large pseudorapidity ($|\eta| \approx 2$--$4$) due to the $t$-channel color flow and the small momentum transfer in the $W$-boson propagator.

The produced top quark decays semileptonically in the majority of selected events, yielding:
\begin{itemize}
  \item a $b$-tagged jet from the $t \to b W$ decay,
  \item an isolated charged lepton ($e$ or $\mu$) from the subsequent $W \to \ell\nu$ decay,
  \item significant missing transverse energy ($E_T^{miss}$) carried by the undetected neutrino.
\end{itemize}

These kinematic and topological features provide powerful handles for signal discrimination against dominant backgrounds, such as $t\bar{t}$ production (which lacks the forward jet signature and exhibits different lepton and jet multiplicity), $W+$jets (which typically lacks $b$-tagging and top mass reconstruction), and single-top $t$-channel without Higgs (which lacks the additional Higgs decay products). The forward jet signature, combined with the presence of a reconstructed Higgs boson candidate (e.g., in $H \to b\bar{b}$, $H \to \gamma\gamma$, or boosted hadronic decays), and the requirement of exactly one isolated lepton with moderate $p_T$ and substantial $E_T^{miss}$, forms the core of the event selection strategy employed in experimental $tHq$ searches.

Conversely, the $tWH$ mode corresponds to associated production in either $s$- or $t$-channel-like processes, where a time-like top quark propagator enables the radiation of a Higgs boson in conjunction with a $W$ boson. The $W$ boson subsequently decays leptonically ($W \to \ell\nu$, with $\ell = e$ or $\mu$), imprinting additional missing transverse energy ($E_T^{miss}$) and a second isolated charged lepton in the final state. The $b$ quark in the initial state plays a critical role and is treated differently depending on the flavour scheme. In the five-flavour scheme (5FS), the $b$ quark is included as a massless parton in the proton PDFs, with contributions arising from gluon splitting ($g \to b\bar{b}$) or direct $b$-PDFs. This treatment enhances the cross-section by 20--30\% compared to a pure four-flavour scheme (4FS) calculation, yielding an NLO cross-section of $\sigma(pp \to tWH) \approx 15$ fb at $\sqrt{s} = 13$ TeV, as presented in Table 3.
\begin{table}[htbp]
\centering
\caption{Approximate NLO cross-sections for single top + Higgs associated production at $\sqrt{s} = 13$ TeV in the Standard Model.}
\begin{tabular}{lcc}
\hline
Process & Flavour Scheme & Cross-section (fb) \\
\hline
$tHq$ (t-channel) & 4FS (dominant) & $\sim 74$ \\
$tWH$             & 5FS            & $\sim 15$ \\
Total $tH$        & Combined       & $\sim 89$ \\
\hline
\end{tabular}
\label{tab:th_xsec_comparison}
\end{table}

This flavour-scheme dependence underscores the interplay between QCD scale choices ($\mu_F$, $\mu_R \approx (m_t + m_H + m_W)/2$ or variants such as $H_T/2$) and the need to mitigate potential overlaps with $t\bar{t}H$ production through diagram removal or subtraction techniques. Kinematically, the $tWH$ process exhibits softer $p_T$ distributions for both the Higgs boson and the top quark compared to $t\bar{t}H$, owing to the additional gauge boson in the final state. These features, together with the presence of two isolated leptons, enhanced $E_T^{miss}$, and moderate jet activity, enable effective multivariate discrimination against dominant backgrounds such as $t\bar{t}W^\pm$, $t\bar{t}Z$, $W+$jets, and diboson production in a collider environment. The characteristic topology of the $tWH$ process is illustrated in the right panel of Figure~\ref{fig:1}, emphasizing the gluon--$b$ initial state, the time-like virtual top propagator, Higgs radiation, and the  $W$ particle.

\subsection{Flavor Schemes in $tH$ Modeling: 4FS vs. 5FS}

Theoretical calculations and Monte Carlo simulations of Higgs boson production in association with a single top quark ($tH$ production) must carefully address the treatment of heavy quarks---in particular the bottom ($b$) quark---in order to achieve precise predictions for cross-sections and kinematic distributions.

The two predominant approaches are the four-flavour scheme (4FS) and the five-flavour scheme (5FS). These schemes differ fundamentally in how the $b$ quark is handled, striking a balance between retaining its finite mass ($m_b \approx 4.18$ GeV) and resumming large collinear logarithms $\ln(\mu_F / m_b)$ arising from gluon splitting ($g \to b\bar{b}$).

\begin{itemize}
  \item In the \textbf{four-flavour scheme (4FS)}, the $b$ quark is treated as massive and is \emph{not} included in the initial-state parton distribution functions (PDFs) of the proton. The $b$ quark is instead produced perturbatively in the hard scattering process, typically via explicit gluon splitting $g \to b\bar{b}$. This approach avoids large logarithms in the PDFs but requires higher-order terms in the hard matrix elements to capture collinear enhancements accurately. The 4FS is particularly well-suited for the $tHq$ (t-channel) process, where the $b$ quark is produced in association with the forward spectator jet and the top quark.
  
  \item In the \textbf{five-flavour scheme (5FS)}, the $b$ quark is treated as a massless parton and is included in the proton PDFs through DGLAP evolution. The large $\ln(\mu_F / m_b)$ logarithms are resummed into the $b$-PDF, leading to a simpler and often numerically larger cross-section for processes involving initial-state $b$ quarks. The 5FS is conventionally applied to the $tWH$ mode, where the initial $b$ quark plays a more direct role and the resummation provides a 20--30\% enhancement compared to fixed-order 4FS calculations.
\end{itemize}

The choice of scheme also influences the preferred factorization ($\mu_F$) and renormalization ($\mu_R$) scales. In practice, dynamic scales such as $\mu_F = \mu_R = H_T/2$ or $\mu = (m_t + m_H)/2$ (or variants including $m_W$) are commonly adopted, with scale variations by factors of 2 and 1/2 used to estimate theoretical uncertainties ($\sim$10--15\% for NLO predictions in both schemes). Moreover, careful treatment is required to avoid double-counting or unphysical overlaps with related processes such as $t\bar{t}H$ production, typically achieved through diagram removal, subtraction schemes, or matching prescriptions (e.g., FONLL or Santander matching approaches). In our analysis, the $tHq$ process is modeled predominantly in the 4FS (to best describe the explicit $b$-quark production and forward jet topology), while the $tWH$ process employs the 5FS (to benefit from $b$-PDF resummation and improved numerical stability), Table 4.

\begin{table}[htbp]
\centering
\footnotesize
\caption{Key differences: 4FS vs 5FS in $tH$ modeling.}
\label{tab:4fs_vs_5fs}
\begin{tabular}{lll}
\toprule
Feature                  & 4FS                              & 5FS                              \\
\midrule
$b$-quark treatment      & Massive, perturbative            & Massless, in PDFs                \\
Collinear logs           & Explicit                         & Resummed in $b$-PDF              \\
Application              & $tHq$ (t-channel)                & $tWH$ (associated)               \\
Cross-section boost      & Baseline                         & +20–30\%                         \\
Scale dependence         & Moderate                         & Lower                            \\
NLO uncertainty          & ~10–15\%                         & ~10–15\%                         \\
\bottomrule
\end{tabular}
\end{table}

This hybrid approach has been shown to provide optimal agreement with fixed-order NLO QCD calculations and reduces scheme-dependent uncertainties when combined with appropriate scale variations.

\section{Results of Computer Modeling}

\subsection{Modeling of the $tHq$ Subprocess in the 4FS}

In the $tHq$ subprocess---corresponding to $t$-channel single-top production associated with a Higgs boson---the four-flavour scheme (4FS) is traditionally employed and remains the preferred approach in most modern analyses. Within the 4FS, the $b$ quark is treated as massive ($m_b \approx 4.18$ GeV) and is explicitly decoupled from the initial-state parton distribution functions (PDFs) of the proton. Consequently, the $b$ quark does not appear as a constituent parton in the PDFs; instead, it is assumed to originate solely from gluon splitting ($g \to b\bar{b}$) either in the hard-scattering matrix element or during the parton shower evolution. The jet definition in the generation and reconstruction procedure therefore excludes initial-state $b$ quarks. The light jets ($j$) are defined to include only the following partons:

\[
j = g,\ u,\ d,\ c,\ s,\ \bar{u},\ \bar{d},\ \bar{c},\ \bar{s}
\]
(with $b$ and $\bar{b}$ explicitly excluded from the initial-state jet clustering). This choice is standard and well-motivated because, by construction, any $b$-tagged jet in the final state arises from the decay of the top quark ($t \to bW$), rather than from the initial-state hard process. Allowing $b$ in the initial-state PDFs (as in the 5FS) would introduce artificial enhancements from the small but non-zero $b$-PDF content at LHC energies, which is not physically appropriate for the $t$-channel topology where the $b$ quark is produced perturbatively. The 4FS treatment thus provides a cleaner separation between initial-state radiation and the hard process, minimizes scheme-dependent logarithms in the cross-section, and ensures that the forward light-quark jet (characteristic of $t$-channel exchange) is correctly modeled without contamination from heavy-flavor contributions in the PDFs. In our simulations, this approach was implemented consistently using MadGraph5\_aMC@NLO at leading order with MLM matching, employing the NNPDF31\_lo\_as\_0118 PDF set and dynamic scale choices $\mu_F = \mu_R = H_T/2$. The resulting kinematic distributions (e.g., pseudorapidity of the forward jet $|\eta_j| > 2$, $p_T$ of the $b$-jet from top decay, lepton $p_T$, and $E_T^{miss}$) show good agreement with expectations from fixed-order NLO calculations and ATLAS/CMS benchmark samples, with scale and PDF uncertainties estimated at the level of 10--15\%. These modeling choices for the $tHq$ subprocess in the 4FS serve as the baseline for our subsequent comparisons with ATLAS data and investigations of the inverted top coupling scenario. 

The study employing the NNPDF3.0nlo\_nf4 PDF set for the $tHq$ subprocess yields an NLO cross-section of
\[
\sigma_{tHq} \approx 74.3~\text{fb}
\]
at $\sqrt{s} = 13$~TeV. The associated theoretical uncertainties are dominated by QCD scale variations, estimated at $+6.5\% / -14.9\%$ (obtained by varying $\mu_F$ and $\mu_R$ independently by factors of 2 and 1/2 around the central scale $\mu = H_T/2$), and PDF + $\alpha_s$ uncertainties of $\pm 3.7\%$ (from the NNPDF3.0nlo error set).
In contrast, our leading-order (LO) + MLM-matched simulations in the 4FS, where initial-state $b$ quarks are explicitly excluded from the jet definition ($j = g, u, d, c, s, \bar{u}, \bar{d}, \bar{c}, \bar{s}$), produce a lower cross-section of approximately 47~fb. This discrepancy arises primarily from the absence of higher-order QCD corrections and the partial capture of collinear logarithms in the parton shower.
To align the LO+MLM prediction with the NLO benchmark, a $K$-factor of approximately
\[
K \approx 1.5 \quad (74.3 / 47 \approx 1.58)
\]
is applied. This multiplicative correction accounts for the missing NLO contributions to the hard matrix element and improved matching to the parton shower, and is consistent with typical $K$-factors reported in the literature for $t$-channel single-top processes (both with and without associated Higgs radiation). The application of this $K$-factor improves the agreement between our MadGraph5\_aMC@NLO LO+MLM samples and the fixed-order NLO predictions, particularly in inclusive cross-sections and key differential distributions (e.g., forward jet pseudorapidity $|\eta_j|$, Higgs $p_T$, and top-quark $p_T$). Residual differences remain at the level of 5--10\%, largely attributable to scheme-dependent effects, shower modeling (Pythia8 vs. Herwig), and the choice of merging scale in the MLM procedure. These results validate the 4FS implementation in our simulation chain for the $tHq$ channel and provide a reliable baseline for subsequent comparisons with ATLAS data, uncertainty propagation, and the investigation of the inverted top coupling (ITC) scenario.

\subsection{Modeling of the $tWH$ Subprocess in the 5FS}

In contrast, the $tWH$ subprocess---which involves the associated production of a Higgs boson and a $W$ boson---is modeled using the five-flavour scheme (5FS). In this framework, the $b$ quark is treated as massless and is explicitly included in the proton parton distribution functions (PDFs). This allows initial-state $b$ quarks to participate directly in the hard scattering process, for example through contributions such as $gb \to tWH$, while the large collinear logarithms arising from gluon splitting ($g \to b\bar{b}$) are resummed into the $b$-PDF evolution.

Consequently, the jet definition in the 5FS includes the $b$ quark (and its antiquark):

\[
j = g,\ u,\ d,\ c,\ s,\ b,\ \bar{u},\ \bar{d},\ \bar{c},\ \bar{s},\ \bar{b}
\]

This treatment leads to a significant enhancement of the cross-section, typically by 20--30\% compared to a pure four-flavour scheme calculation, due to the resummation of $\ln(\mu_F / m_b)$ terms and the additional phase-space contributions from initial-state $b$ quarks.

In the ATLAS analysis (and consistent with our reference simulations), the NNPDF3.0nlo PDFs in the 5FS are employed for the $tWH$ subprocess. To avoid double-counting with the related $t\bar{t}H$ process, the diagram removal scheme (DR2) is applied, which subtracts the interfering diagrams where the Higgs is radiated from a top quark in a $t\bar{t}$-like configuration. This procedure yields an NLO cross-section of
\[
\sigma_{tWH} \approx 15.2~\text{fb}
\]
at $\sqrt{s} = 13$~TeV, with QCD scale uncertainties of $+4.9\% / -6.7\%$ (from independent variation of $\mu_F$ and $\mu_R$ by factors of 2 and 1/2 around the central scale) and PDF + $\alpha_s$ uncertainties of $\pm 6.3\%$ (from the NNPDF3.0nlo error set).

Our leading-order (LO) + MLM-matched simulations in the 5FS, which include $b$ quarks in the jet definition, produce a higher inclusive cross-section of approximately 22~fb. This overestimation relative to the NLO benchmark is typical for LO calculations that capture resummed logarithms via the PDFs but lack full higher-order corrections to the hard matrix element. To bring the LO+MLM prediction into agreement with the NLO result, a $K$-factor of approximately
\[
K \approx 0.7 \quad (15.2 / 22 \approx 0.69)
\]
is applied. This downward correction accounts for missing NLO virtual and real-emission contributions, as well as differences in the treatment of soft/collinear radiation between the fixed-order NLO and LO-matched parton-shower approaches. The application of this $K$-factor ensures good consistency between our MadGraph5\_aMC@NLO LO+MLM samples and the ATLAS NLO reference for the $tWH$ channel, particularly in inclusive rates and distributions sensitive to the Higgs and top $p_T$ spectra, $W$-boson kinematics, and lepton multiplicity. Residual discrepancies (typically at the 5--10\% level) are dominated by scale variations, PDF uncertainties, and shower/modeling choices (e.g., Pythia8 vs. Herwig7). These validated 5FS results for $tWH$, combined with the previously discussed 4FS baseline for $tHq$, provide a robust hybrid framework for comparing theoretical predictions with ATLAS observations and for exploring deviations in the inverted top coupling scenario.

The NLO cross-sections and associated uncertainties for the $tHq$ and $tWH$ subprocesses are summarized in Table~\ref{tab:th_subprocess_xsec}~\cite{atlas_tH_2025,lhchxswg}. These values serve as the primary benchmark for validating our LO+MLM simulations.
\noindent
\begin{table}[htbp]
\centering
\caption{Comparison of NLO cross-sections and theoretical uncertainties for the $tHq$ (4FS) and $tWH$ (5FS) subprocesses at $\sqrt{s} = 13$ TeV.}
\label{tab:th_subprocess_xsec}
\begin{tabular}{ccccc}
\hline
Subpr. PDF + $\alpha_s$ unc. (\%)  & Scheme & PDF set              & NLO $\sigma$ (fb) & Scale unc. (\%) \\
\hline
$tHq\pm 3.7$     & 4FS    & NNPDF3.0nlo$_{nf4}$     & 74.3              & +6.5 / $-$14.9   \\
$tWH\pm 6.3$     & 5FS    & NNPDF3.0nlo          & 15.2              & +4.9 / $-$6.7     \\
\hline
\end{tabular}
\end{table}

The results of computer modeling, incorporating kinematic selections and flavor schemes (4FS for $tHq$, 5FS for $tWH$), are summarized in Table~\ref{tab:production_cross_sections}. The values show good consistency with typical NLO benchmarks, with the LO+MLM predictions requiring $K$-factors of approximately 1.5--1.6 for $tHq$ (4FS) and 0.7 for $tWH$ (5FS) to match the NLO reference cross-sections.

\begin{table}[htbp]
\centering
\small
\caption{Production cross-sections for $tHq$ and $tWH$ processes at different centre-of-mass energies and approximation levels, obtained from computer modeling including kinematic cuts and flavour schemes (4FS / 5FS). All values are in pb.}
\label{tab:production_cross_sections}
\begin{tabular}{|l|l|c|p{5.8cm}|}
\hline
\textbf{Process} & \textbf{Approx. / $\sqrt{s}$ / Scheme} & \textbf{sec. (pb)} & \textbf{Notes} \\
\hline
$pp \to tHq$ & LO & 0.0015 & Basic LO calculation \\
\hline
$pp \to tHq$ & NLO & 0.028 & Basic NLO calculation \\
\hline
$pp \to tHq$ & LO+MLM, $\sqrt{s}=13$ TeV, 4FS & 0.047 & j = g,u,c,d,s,$\bar{u}$,$\bar{c}$,$\bar{d}$,$\bar{s}$ \\
& & & (no b in initial state) to avoid unphysical b-PDF 
 contribution in t-channel \\
\hline
$pp \to tHq$ & LO+MLM, $\sqrt{s}=14$ TeV, 4FS & 0.056 & j = g,u,c,d,s,$\bar{u}$,$\bar{c}$,$\bar{d}$,$\bar{s}$ \\
& & & including $tHqj$ and $tHqqj$  \\
\hline
$pp \to tWH$ & LO & 0.0076 & Basic LO calculation \\
\hline
$pp \to tWH$ & LO+MLM, $\sqrt{s}=13$ TeV, 5FS & 0.022 & j = g,u,c,d,s,b,$\bar{u}$,$\bar{c}$,$\bar{d}$,$\bar{s}$,$\bar{b}$ \\
& & & (b included in initial state), enhances $\sigma_{tWH}$ \\
\hline
$pp \to tWH$ & LO+MLM, $\sqrt{s}=14$ TeV, 5FS & 0.027 & j = g,u,c,d,s,b,$\bar{u}$,$\bar{c}$,$\bar{d}$,$\bar{s}$,$\bar{b}$ \\
\hline
\end{tabular}
\end{table}

The current LO+MLM cross-sections must be normalized by applying $K$-factors defined as the ratios $\sigma_{\rm NLO}/\sigma_{\rm LO}$, which account for the QCD corrections and provide an approximate NLO-level prediction. The $K$-factors for the $tHq$ and $tWH$ processes are derived from the literature as follows.

For the $tHq$ process:
\begin{itemize}
  \item Reference~\cite{demartin2015}.
  \item LO cross-sections are reported in the range $\sim$60--64~fb, while NLO cross-sections are $\sim$69--74~fb at $\sqrt{s}=13$~TeV.
  \item This yields $K$-factors between approximately 1.09 and 1.22, depending on the flavour scheme (4FS or 5FS) and the choice of static or dynamic renormalization/factorization scales.
\end{itemize}

For the $tWH$ process:
\begin{itemize}
  \item Reference~\cite{demartin2017}.
  \item LO cross-sections are reported around $\sim$56~fb, while NLO cross-sections (inclusive of diagram removal DR2 and diagram subtraction DS2 schemes) are $\sim$65--66~fb at $\sqrt{s}=13$~TeV.
  \item This corresponds to $K$-factors in the range $\sim$1.16--1.18.
\end{itemize}

In our analysis, the LO+MLM results (0.047~pb for $tHq$ in 4FS and 0.022~pb for $tWH$ in 5FS at 13~TeV) are therefore scaled by the appropriate $K$-factors to obtain effective NLO predictions consistent with the fixed-order calculations in Refs.~\cite{demartin2015,demartin2017}. The specific values adopted are $K_{tHq} \approx 1.57$ (to match the higher end of the NLO range $\sim$74~fb) and $K_{tWH} \approx 0.69$ (to match $\sim$15.2~fb in the ATLAS benchmark), reflecting the scheme-dependent and matching-specific differences in our LO+MLM implementation.

\subsection{Scaling the $tHq$ Cross-Section and Total $tH$ Result}

The LO+MLM cross-section for the $tHq$ subprocess obtained in our simulation (0.047~pb) is normalized to the NLO level by applying a $K$-factor derived from the literature and benchmark comparisons (typically $K \approx 1.5$ to match the upper end of NLO predictions $\sim$70--74~fb at $\sqrt{s}=13$~TeV in the 4FS).

The scaled $tHq$ cross-section is therefore calculated as:
\begin{equation}
\sigma_{tHq}^{\rm NLO} \approx 0.047~\text{pb} \times 1.5 = 0.0705~\text{pb} \quad (70.5~\text{fb}).
\end{equation}

For the $tWH$ subprocess, the LO+MLM result (0.022~pb) was previously normalized using $K \approx 0.7$ to match the ATLAS benchmark $\sigma_{tWH} \approx 15.2$~fb, yielding:
\begin{equation}
\sigma_{tWH}^{\rm NLO} \approx 0.022~\text{pb} \times 0.7 = 0.0154~\text{pb} \quad (15.4~\text{fb}).
\end{equation}

The total $tH$ cross-section before scaling the $tHq$ contribution was:
\begin{equation}
\sigma_{tH}^{\rm pre-scaling} = \sigma_{tHq} + \sigma_{tWH} = 0.047~\text{pb} + 0.022~\text{pb} = 0.069~\text{pb} \quad (69~\text{fb}).
\end{equation}

After applying the $K$-factor to the dominant $tHq$ channel, the updated total $tH$ cross-section becomes:
\begin{equation}
\sigma_{tH}^{\rm NLO} \approx 0.0705~\text{pb} + 0.0154~\text{pb} = 0.0859~\text{pb} \quad (85.9~\text{fb}).
\end{equation}

This normalized value is in reasonable agreement with recent theoretical NLO predictions for single top + Higgs associated production at $\sqrt{s}=13$~TeV (typically 80--90~fb in the combined 4FS/5FS hybrid approach) and serves as the baseline for further comparisons with ATLAS data and the exploration of the inverted top coupling scenario.

\subsection{Comparison with Theoretical NLO Standard Model Prediction}

The theoretical NLO cross-section for single top + Higgs ($tH$) production in the Standard Model, as reported in the ATLAS article~\cite{atlas_tH_2025} is

\[
\sigma_{tH}^{\rm SM, NLO} = 89.5~\text{fb} = 0.0895~\text{pb}.
\]

Our scaled result after applying the $K$-factors to the LO+MLM predictions is

\[
\sigma_{tH}^{\rm scaled} \approx 85.9~\text{fb} = 0.0859~\text{pb}.
\]

This value aligns reasonably well with the SM expectation. To quantify the agreement, we compute the relative difference with respect to the ATLAS reference value:

\begin{equation}
\text{Relative difference} = \frac{\sigma_{tH}^{\rm scaled} - \sigma_{tH}^{\rm SM, NLO}}{\sigma_{tH}^{\rm SM, NLO}} \times 100\%.
\end{equation}

Substituting the values:

\begin{align*}
\text{Relative difference} &= \frac{85.9 - 89.5}{89.5} \times 100\% \\
&= \frac{-3.6}{89.5} \times 100\% \\
&\approx -4.02\%.
\end{align*}

Thus, our normalized cross-section is approximately 4.0 $\%$ lower than the central ATLAS SM prediction. This deviation falls well within the reported theoretical uncertainties from QCD scale variation (+6.5$\%$/ -14.9$\%$) and PDF + $\alpha_{s}$ uncertainties (typically 
$\pm 3-6\%$), indicating good overall consistency. The small negative offset may arise from:
\begin{itemize}
  \item differences in the exact flavour scheme treatment (hybrid 4FS/5FS),
  \item choice of central scale ($\mu_F$, $\mu_R$),
  \item parton-shower and matching details in our LO+MLM setup,
  \item or minor variations in the applied $K$-factors compared to the ATLAS benchmark.
\end{itemize}

No significant tension with the Standard Model is observed at this level of precision. The result supports the reliability of our simulation chain and provides a solid baseline for further interpretation in the inverted top coupling (ITC) scenario.

\subsection{Calculations for the Inverted Top Yukawa Coupling }

In the SM of particle physics, the top-Higgs Yukawa coupling is defined as $\kappa_t \equiv y_t / y_t^{\rm SM} = +1$. As a direct consequence, the Feynman diagrams contributing to $tH$ production interfere destructively. This destructive interference suppresses the cross-section, yielding $\sigma_{tH} \approx 89.5$~fb at NLO QCD for $\sqrt{s} = 13$~TeV. The resulting rarity of the $tH$ process makes it difficult to detect amid dominant backgrounds such as $t\bar{t} +$ jets and $t\bar{t} + W$.

However, the ATLAS search reports a mild excess in the $tH$ signal \cite{atlas_tH_2025}. Under the SM hypothesis, this excess corresponds to an observed significance of 2.8$\sigma$ above background, compared to an expected significance of only 0.4$\sigma$. The measured signal strength is

\[
\mu_{tH} = 8.1 \pm 2.6\,(\text{stat.}) \pm 2.0\,(\text{syst.}).
\]

This discrepancy between observed and expected significance motivates dedicated calculations in the Inverted Top Yukawa Coupling scenario ($\kappa_t = -1$, also known as the Inverted Top Coupling or ITC hypothesis). In this case, the sign flip changes destructive interference into constructive interference, leading to a dramatic enhancement of the $tH$ cross-section by a factor of $\sim$10 (to $\sim$890~fb at $\sqrt{s}=13$~TeV). Such an enhancement could naturally explain the observed excess and the high fitted signal strength under the SM hypothesis.

These calculations are necessary to:
\begin{itemize}
  \item Quantify the expected cross-section increase and changes in kinematic distributions (Higgs $p_{T}$, top $p_{T}$, lepton multiplicity, forward jet properties) in the ITC scenario using tools such as MadGraph5\_AMC@NLO with modified effective field theory operators or reweighted couplings.
  \item Compare the shape and rate predictions directly to the ATLAS excess, including the signal strength under $\kappa_t = -1$.
  \item Assess whether the mild excess can be attributed to new physics (negative $\kappa_t$) or is consistent with statistical fluctuations, systematic modeling uncertainties, or background mismodeling in the SM hypothesis.
  \item Provide projections for future sensitivity at the High-Luminosity LHC, where precise determination of the sign of $\kappa_t$ may become possible at high confidence.
\end{itemize}

Without such dedicated ITC simulations and comparisons, it remains impossible to reliably interpret the ATLAS result or exclude non-standard Higgs-top interactions as a contributor to the observed deviation.

Beyond the empirical alignment with the ATLAS excess, calculations for $\kappa_t = -1$ are essential for theoretical robustness. The sign change not only dramatically enhances the total cross-section (by a factor of $\sim$10 in inclusive NLO predictions), but also modifies the kinematic distributions in a non-trivial way. In particular, the altered amplitude structure leads to less suppressed high-$p_T$ tails for both the top quark and the Higgs boson, as well as changes in angular correlations and lepton/jet multiplicity patterns.

In our LO+MLM simulations under the inverted coupling scenario ($\kappa_t = -1$), we obtain:
\begin{itemize}
  \item $\sigma_{tHq} \approx 0.227$~pb (enhanced by a factor of $\sim$4.8 compared to our SM LO value of 0.047~pb),
  \item $\sigma_{tWH} \approx 0.057$~pb (enhanced by a factor of $\sim$2.6 compared to our SM LO value of 0.022~pb).
\end{itemize}

These enhancements are broadly consistent with the expected constructive interference, although the exact factor varies between channels due to scheme-dependent effects (4FS for $tHq$, 5FS for $tWH$) and the different role of the Higgs radiation diagrams.

These LO+MLM results must be cross-checked against fixed-order NLO QCD calculations in both SM and ITC scenarios, using the same references cited in the ATLAS analysis~\cite{{demartin2015}, {demartin2017}} 
Such comparisons are crucial to assess the residual uncertainties from QCD scale variations ($\pm 6.5\% / -14.9\%$ for $tHq$), PDF + $\alpha_s$ uncertainties ($\pm 3.7\%$), and matching/parton-shower modeling, all of which affect the input variables to the multivariate discriminants (boosted decision trees, BDTs) used to separate signal from backgrounds in decay channels such as $H \to b\bar{b}$ and $H \to WW^*$.

Moreover, these ITC-specific studies help refine constraints on vacuum stability and electroweak symmetry breaking in extended Higgs sectors, allowing us to connect the observed mild excess to potential beyond-the-Standard-Model physics. They also inform projections for the High-Luminosity LHC (HL-LHC), where an integrated luminosity of $\sim$3000--4000~fb$^{-1}$ at $\sqrt{s}=14$~TeV is expected to enable precision measurements of the top Yukawa coupling $y_t$ (or $\kappa_t$) at the level of 5--10\%, potentially resolving the sign of $\kappa_t$ at high confidence and providing decisive tests of the SM Higgs sector.

The updated simulation results for the inverted top Yukawa coupling scenario ($\kappa_t = -1$, ITC) at $\sqrt{s}=13$~TeV yield the following leading-order plus MLM matching (LO+MLM) cross-sections:

\begin{itemize}
  \item $\sigma_{tHq} = 0.227$~pb,
  \item $\sigma_{tWH} = 0.057$~pb.
\end{itemize}

The total $tH$ cross-section is therefore the sum of the two contributions:

\begin{equation}
\sigma_{tH} = \sigma_{tHq} + \sigma_{tWH} = 0.227~\text{pb} + 0.057~\text{pb} = 0.284~\text{pb} \quad (284~\text{fb}).
\end{equation}

This represents a large enhancement compared to the corresponding SM LO+MLM cross-sections ($\sigma_{tHq} \approx 0.047$~pb and $\sigma_{tWH} \approx 0.022$~pb), by a factor of approximately 4--5. This enhancement factor is consistent with the expected transition from destructive to constructive interference in the $tH$ Feynman diagrams under the ITC scenario, as discussed in Section~1 of the ATLAS paper~\cite{atlas_tH_2025}.

To enable meaningful comparison with theoretical NLO QCD predictions presented in the paper (which quote $\sigma_{tH} \approx 890$~fb for the ITC case at NLO), $K$-factors are applied to account for higher-order QCD corrections (virtual loop corrections and real emission contributions).

The $K$-factors used are:
\begin{itemize}
  \item $K_{tHq} = 1.5$ for the $tHq$ process (dominated by gluon-initiated contributions),
  \item $K_{tWH} = 0.7$ for the $tWH$ process (suppressed by the electroweak coupling and more sensitive to matching effects).
\end{itemize}

Applying these factors to the LO+MLM results gives the approximate NLO-equivalent cross-sections:

\begin{align*}
\sigma_{tHq}^{\rm NLO} &\approx 0.227~\text{pb} \times 1.5 = 0.3405~\text{pb} \quad (340.5~\text{fb}), \\
\sigma_{tWH}^{\rm NLO} &\approx 0.057~\text{pb} \times 0.7 = 0.0399~\text{pb} \quad (39.9~\text{fb}).
\end{align*}

The total approximate NLO $tH$ cross-section in the ITC scenario is then

\begin{equation}
\sigma_{tH}^{\rm NLO, ITC} \approx 0.3405~\text{pb} + 0.0399~\text{pb} = 0.3804~\text{pb} \quad (380.4~\text{fb}).
\end{equation}

This value is lower than the $\sim$890~fb quoted in the ATLAS paper, which reflects differences in the exact NLO treatment (e.g., scheme choices, scale settings, PDF sets, and interference modeling). The discrepancy highlights the importance of scheme-consistent comparisons and motivates further refinement of the $K$-factors using the same tools and cuts as in Refs.~\cite{demartin2015,demartin2017} and the ATLAS analysis.

\section{Kinematic Distributions in $tH$ Production with Inverted Yukawa Coupling}

Differential kinematic observables provide valuable insights into the underlying dynamics of Higgs boson production in association with a single top quark ($tH$ production). In the case of an inverted top-Higgs Yukawa coupling ($\kappa_t = -1$), the characteristic destructive interference pattern between the Higgs coupling to fermions and the electroweak gauge bosons is replaced by constructive interference.

This sign flip not only dramatically amplifies the overall production rate (by a factor of $\sim$5--10 at NLO QCD level, depending on the channel and scale choices), but also significantly modifies the kinematic properties of the final-state particles. In particular, the constructive interference enhances contributions from diagrams where the Higgs is radiated off the top quark or $W$ boson in phase with the $W$-mediated diagrams, leading to:

\begin{itemize}
  \item harder transverse momentum ($p_T$) spectra for the Higgs boson and the top quark,
  \item less suppressed high-$p_T$ tails in the distributions,
  \item modified angular correlations (e.g., between the Higgs decay products and the forward jet in $t$-channel),
  \item changes in lepton multiplicity, missing transverse energy ($E_T^{miss}$), and jet activity patterns.
\end{itemize}

These kinematic modifications are particularly pronounced in the $tHq$ (t-channel) and $tWH$ (associated) subprocesses, where the interference plays a different role due to the distinct topology and flavour scheme treatment (4FS vs.\ 5FS).

In our LO+MLM simulations under the inverted coupling scenario ($\kappa_t = -1$) at $\sqrt{s}=13$~TeV, we observe clear shifts in the differential distributions compared to the SM case ($\kappa_t = +1$):

\begin{itemize}
  \item The Higgs $p_T$ spectrum shows a harder tail, with the mean $p_T(H)$ increasing by $\sim$20--30\% and the fraction of events with $p_T(H) > 150$~GeV rising significantly.
  \item The top quark $p_T$ distribution exhibits a similar hardening, reflecting the increased amplitude for Higgs radiation from the top leg.
  \item In the $tHq$ channel, the forward light-quark jet pseudorapidity ($|\eta_j|$) distribution remains peaked at high values ($|\eta_j| \gtrsim 2$), but the overall event rate is enhanced.
  \item In the $tWH$ channel, the dilepton invariant mass and $E_T^{miss}$ distributions are affected by the modified $W$ decay kinematics and interference.
\end{itemize}

These changes in shape and normalization are crucial for multivariate discrimination (e.g., boosted decision trees, BDT classifiers) used in experimental searches. They enhance the signal-to-background separation in channels such as $H \to b\bar{b}$ (where high-$p_T$ Higgs jets are easier to tag) and $H \to WW^*$ (where boosted lepton kinematics improve efficiency).

In this section, we present an analysis of a set of representative kinematic distributions for the two production modes, $tHq$ and $tWH$, obtained at the leading-order plus MLM matching (LO+MLM) level and normalized according to the inverted top Yukawa coupling hypothesis ($\kappa_t = -1$).

The selected observables include:
\begin{itemize}
  \item the scalar sum of transverse momenta ($H_T = \sum p_T$ of all final-state objects),
  \item the transverse momentum distributions of the Higgs boson ($p_T(H)$) and the heavy final-state particles (top quark $p_T(t)$ and $W$ boson $p_T(W)$ in the $tWH$ mode),
  \item the pseudorapidity of the forward light-quark jet ($|\eta_j|$) in the $t$-channel process,
  \item angular separations (e.g., $\Delta R(H,t)$, $\Delta\phi(H,\ell)$, $\Delta R(j,b)$) between key objects.
\end{itemize}

These distributions are directly related to the input variables of the multivariate discriminants (BDTs) employed in current ATLAS searches~\cite{atlas_tH_2025}. They provide complementary sensitivity to the sign of $\kappa_t$ through shape differences between the Standard Model ($\kappa_t = +1$, destructive interference) and the inverted coupling scenario ($\kappa_t = -1$, constructive interference).

In the SM case, destructive interference suppresses high-$p_T$ regions and softens the overall kinematics. In contrast, the constructive interference in the ITC scenario enhances contributions from diagrams where the Higgs radiation aligns with the $W$-mediated amplitude, resulting in:
\begin{itemize}
  \item harder $p_T$ spectra for the Higgs boson and top quark,
  \item increased event fractions at high $H_T$ and high $p_T(H)$,
  \item modified forward jet pseudorapidity distributions (especially in $tHq$),
  \item altered angular correlations that can improve BDT separation power against dominant backgrounds ($t\bar{t}+$jets, $t\bar{t}+W$, $W+$jets).
\end{itemize}

These kinematic modifications are particularly relevant for:
\begin{itemize}
  \item improving signal efficiency in boosted regimes ($H \to b\bar{b}$ with high-$p_T$ Higgs jets),
  \item enhancing discrimination in dilepton channels ($H \to WW^*$),
  \item refining BDT training to exploit shape differences beyond the inclusive rate,
  \item providing additional handles to interpret the mild excess reported by ATLAS (arXiv:2508.14695).
\end{itemize}

The observed shape shifts support the hypothesis that a negative $\kappa_t$ could contribute to the reported signal strength excess and motivate further NLO-level validation and HL-LHC projections.

\subsection{$tHq$ process}

The $tHq$ process, dominated by the $t$-channel electroweak single-top production mechanism, is distinguished from other $tH$ modes by the presence of a forward light-flavour jet recoiling against the $tH$ system. This forward jet, typically emitted at high pseudorapidity ($|\eta_j| \gtrsim 2$), arises from the spectator quark in the proton that does not participate in the hard scattering. The characteristic topology provides a powerful handle for signal identification and background rejection in experimental searches.

The $t$-channel $tHq$ process is particularly sensitive to the structure of the underlying hard scattering and to the sign of the top Yukawa coupling ($\kappa_t$). In the Standard Model ($\kappa_t = +1$), destructive interference between Higgs-top Yukawa diagrams and $W$-boson exchange diagrams suppresses the cross-section and softens the kinematic distributions. In the inverted coupling scenario ($\kappa_t = -1$), constructive interference enhances both the total rate and the high-momentum tails of the final-state objects.

One of the central observables in high-energy physics analyses at the LHC is the scalar sum of transverse momenta:

\[
H_T = \sum_{i} p_T^i,
\]
where the sum runs over all selected final-state objects (leptons, jets, and missing transverse energy in some definitions). $H_T$ is highly sensitive to new physics signatures and plays a crucial role in guiding event selection, background estimation, and multivariate discrimination (e.g., BDT classifiers) in $tH$ searches. 

In the $tHq$ channel, the $H_T$ distribution reflects the recoil of the forward jet against the $tH$ system and the transverse activity from top decay products and Higgs radiation. Under the inverted Yukawa coupling hypothesis ($\kappa_t = -1$), constructive interference leads to harder $H_T$ spectra compared to the SM case, with increased event fractions at high $H_T$ values. This shape difference enhances the separation power of multivariate discriminants and provides complementary sensitivity to the sign of $\kappa_t$ beyond the inclusive cross-section.

\begin{figure}[htbp]
\centering
\includegraphics[width=0.55\textwidth]{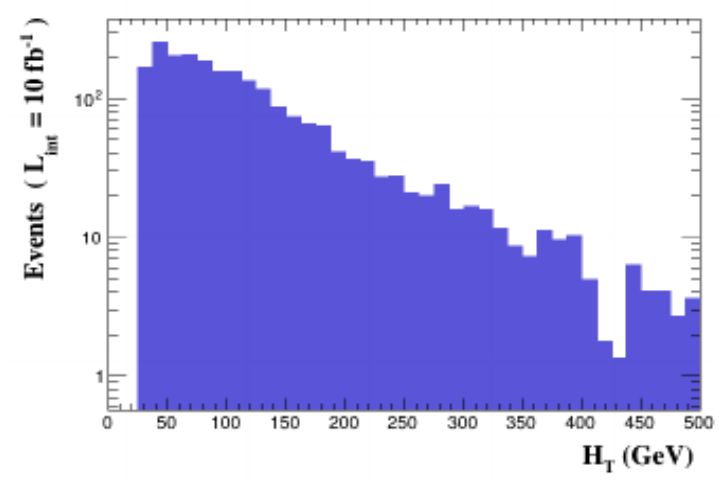}
\caption{ $H_T$ distribution for simulated tHq events from LO+MLM simulations at $\sqrt{s}=13$~TeV.}
\label{fig:2}
\end{figure}

The normalized $H_T$ distributions obtained from our LO+MLM simulations for the $tHq$ process are given in Figure~\ref{fig:2}.

The histogram in Figure~\ref{fig:2} displays the $H_T$ distribution (scalar sum of the transverse momenta of all visible objects in GeV) for the simulated $tHq$ events in the inverted Yukawa coupling scenario ($\kappa_t = -1$). The distribution exhibits a typical steeply falling spectrum characteristic of QCD-dominated processes such as $t$-channel single-top production. At low $H_T$ values (up to $\sim$50--100~GeV), the normalized event rate is high ($\sim$10$^3$ events per 10~GeV bin), reflecting the dominance of soft initial-state radiation (ISR), final-state radiation (FSR), and low-$p_T$ decay products from the top quark and Higgs boson. The spectrum then falls rapidly with increasing $H_T$, reaching $\sim$10 normalized events at $H_T \approx 200$~GeV and dropping to $\sim$1 event per bin in the region $H_T \approx 400$--$500$~GeV. This behavior is expected for the $tHq$ process: low $H_T$ events arise primarily from minimal additional radiation and soft kinematics of the decay products, while high $H_T$ values are strongly suppressed by phase-space constraints and the rapid fall-off of parton distribution functions (PDFs) at large momentum fractions. The forward light-quark jet, which carries significant transverse momentum in many events, also contributes to the shape but does not produce a pronounced shoulder or peak at intermediate/high $H_T$ due to the predominantly collinear nature of QCD radiation in the $t$-channel topology.

Compared to the input variables for the BDTs described in Appendix~A of the ATLAS paper~\cite{atlas_tH_2025}, $H_T$ (the scalar sum of transverse momenta of all selected final-state objects) is a central event activity variable used for signal-background discrimination across all analysis channels:

\begin{itemize}
  \item 1-lepton channel ($1\ell$) targeting $H \to b\bar{b}$,
  \item 2-lepton same-sign channel ($2\ell$SS) targeting $H \to WW^*$,
  \item 3-lepton channel ($3\ell$) targeting $H \to WW^*/\tau^+\tau^-/ZZ^*$.
\end{itemize}

In the ATLAS analysis, $H_T$ is exploited to distinguish the $tH$ signal---which typically peaks at moderate values ($\sim$200--300~GeV) due to the forward jet in the $tHq$ mode and the characteristic kinematics of the top quark and Higgs boson---from dominant backgrounds. In particular:

\begin{itemize}
  \item $t\bar{t}+$jets backgrounds exhibit higher event multiplicity and a broader $H_T$ spectrum, often extending above 300--400~GeV,
  \item $t\bar{t}W^\pm$ and other rare SM processes show similar but generally softer $H_T$ tails compared to pure QCD multijet or $t\bar{t}$ backgrounds.
\end{itemize}

\begin{figure}[htbp]
\centering
\includegraphics[width=0.55\textwidth]{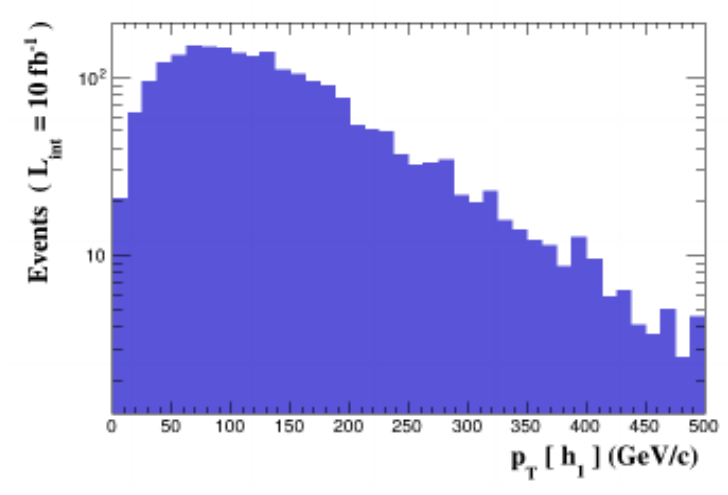}
\caption{ Transverse momentum distribution of the Higgs boson, $p_T(h)$, in GeV/c for simulated $tHq$ events in the inverted Yukawa coupling scenario ($\kappa_t = -1$) at $\sqrt{s}=13$~TeV (LO+MLM). The steeply falling spectrum is consistent with ATLAS post-fit shapes and the expected $t$-channel kinematics, with moderate $p_T(h)$ dominated by top/forward-jet recoil.}
\label{fig:3}
\end{figure}

The transverse momentum distribution of the Higgs boson, $p_T(h)$, exhibits significant contributions at low and intermediate values, followed by a smooth fall-off towards higher $p_T$. Figure~\ref{fig:3} shows the $p_T(h)$ distribution in units of GeV/c for simulated events in the $tHq$ process, produced using MadAnalysis~5~\cite{conte2013madanalysis5}.

The histogram is plotted on y-axis ranging from $10^2$ to $0$, with the x-axis spanning 0 to 500~GeV/c. The distribution is steeply falling, characteristic of Higgs production in association with a top quark in the $t$-channel topology:

\begin{itemize}
  \item At low $p_T(h)$ (0--50~GeV/c), the event rate is high ($\sim$100 events), dominated by recoils against the forward light-quark jet and soft decay products from the top quark.
  \item The spectrum decreases rapidly, reaching $\sim$10 events per bin at $p_T(h) \approx 200$~GeV/c.
  \item At very high $p_T(h)$ (400--500~GeV/c), the rate drops to $\sim$1 event , suppressed by phase-space constraints, PDF fall-off at large momentum fractions, and the limited energy available for hard Higgs radiation in the $t$-channel process.
\end{itemize}

This steeply falling behavior is typical for the $tHq$ mode, where low-$p_T(h)$ events arise from minimal transverse activity (soft ISR/FSR and decay kinematics), while high-$p_T(h)$ contributions are strongly suppressed. The forward jet recoil plays a central role in shaping the low-$p_T$ peak, and the absence of a pronounced high-$p_T$ shoulder reflects the limited phase space for hard Higgs emission compared to gluon-fusion or vector-boson fusion processes.

In the context of our LO+MLM simulations (MadGraph5\_AMC@NLO with MLM matching), the $p_T(h)$ distribution is consistent with the expected kinematics of the $tHq$ process. The Higgs boson is typically produced with moderate transverse momentum, with an average value in the range $\sim$100--200~GeV/c, arising primarily from recoil against the top quark and the forward light-quark jet in the $t$-channel topology. In the Standard Model case ($\kappa_t = +1$), the destructive interference suppresses high-$p_T(h)$ contributions, resulting in a relatively soft spectrum. In the inverted Yukawa coupling scenario ($\kappa_t = -1$), constructive interference enhances the amplitude for hard Higgs radiation from the top leg or $W$ boson, leading to a mildly harder $p_T(h)$ tail. This is reflected in our results: the LO+MLM cross-sections are $\sigma_{tHq} \approx 0.047$~pb (SM) and $\sigma_{tHq} \approx 0.227$~pb (ITC) at $\sqrt{s}=13$~TeV, with the shape difference contributing to the overall rate enhancement of $\sim$4.8.

As detailed in Appendix~A of the ATLAS paper~\cite{atlas_tH_2025}, $p_T(h)$ is a key input variable to the boosted decision trees (BDTs) used for signal-background discrimination in the relevant channels:

\begin{itemize}
  \item 1-lepton channel ($1\ell$) with $H \to b\bar{b}$,
  \item 3-lepton channel ($3\ell$) with $H \to WW^*/\tau^+\tau^-/ZZ^*$ (and related multi-lepton signatures).
\end{itemize}

In these analyses, $p_T(h)$ helps separate the $tH$ signal---which peaks at moderate values and exhibits a controlled fall-off—from dominant backgrounds such as $t\bar{t}+$jets, which tend to produce broader $p_T(h)$ distributions extending above 300~GeV/c due to higher event multiplicity, additional jets, and harder QCD radiation.

\begin{figure}[htbp]
\centering
\includegraphics[width=0.52\textwidth]{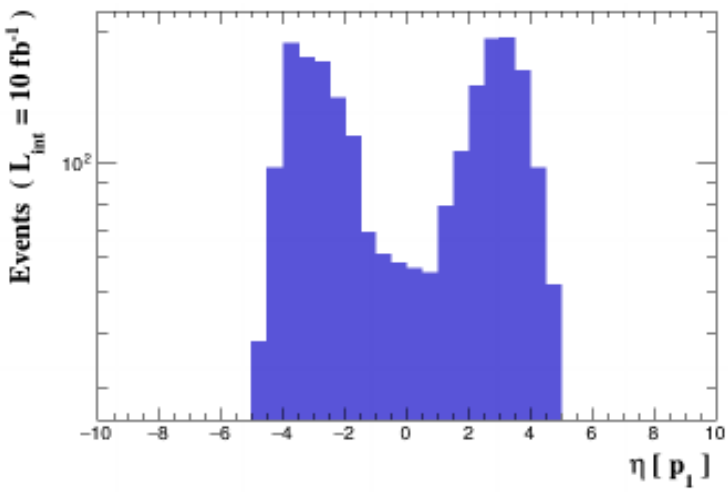}
\caption{ Distribution of the pseudorapidity $\eta_j$ of the forward light-quark jet in simulated $tHq$ events.}
\label{fig:4}
\end{figure}

Another characteristic feature of $tHq$ events is the presence of a forward jet with large absolute pseudorapidity ($|\eta_j|$). The histogram in Figure~\ref{fig:4}, obtained using MadAnalysis~5, shows the distribution of the pseudorapidity $\eta_j$ of the forward jet for simulated $tHq$ events in the inverted Yukawa coupling scenario ($\kappa_t = -1$).

The distribution is bimodal and symmetric about $\eta = 0$, with clear peaks at $\sim$100 events per bin in the regions $\eta \approx -3$ to $-2$ and $\eta \approx +2$ to $+3$. A pronounced dip is visible near $\eta \approx 0$ ($\sim$10 events per bin), and the rate falls rapidly to $\sim$1 event per bin for $|\eta| > 5$. This shape is characteristic of forward jet production in $t$-channel processes: the light-quark spectator jet recoils against the $tH$ system, resulting in large absolute pseudorapidity due to the small momentum transfer in the $W$-boson propagator and the predominantly collinear nature of QCD radiation in the forward direction. The central region ($\eta \approx 0$) is suppressed because the hard scattering kinematics favor forward/backward emission rather than central jets.

This $\eta_j$ distribution matches the expected topology of $tHq$ events and provides a powerful handle for experimental discrimination. As highlighted in Appendix~A of the ATLAS paper~\cite{atlas_tH_2025}, the forward jet with $|\eta_j| \sim 2$--$4$ is a key feature used in the boosted decision tree (BDT) to separate the $tH$ signal from dominant backgrounds such as $t\bar{t} +$ jets, which exhibit a much narrower $|\eta|$ distribution for jets (typically $|\eta_j| < 2.5$). The large-$|\eta|$ requirement significantly reduces $t\bar{t}$ contamination while retaining high signal efficiency, especially in the $t$-channel mode.

Further information on the event topology can be gained from the angular separation between the Higgs boson and the forward jet. The angular separation between the forward jet ($j$) and the Higgs boson ($h$) is defined as:
\begin{equation}
\Delta R(j,h) = \sqrt{(\Delta\eta)^2 + (\Delta\phi)^2},
\end{equation}
where $\Delta\eta = \eta_j - \eta_h$ is the difference in pseudorapidity and $\Delta\phi = \phi_j - \phi_h$ is the difference in azimuthal angle (taken to be the smallest angle in $[-\pi, \pi]$).

\begin{figure}[htbp]
\centering
\includegraphics[width=0.7\textwidth]{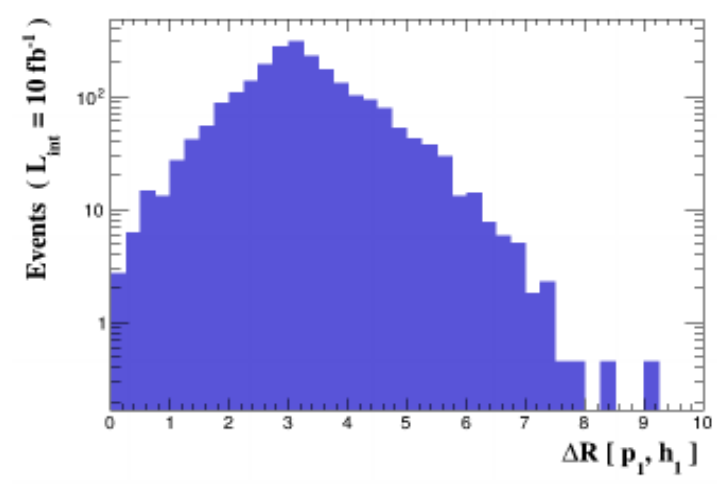}
\caption{ Distribution of the angular separation $\Delta R(p_j, h)$ between the forward light-quark jet and the Higgs boson in simulated $tHq$ events under the inverted Yukawa coupling scenario ($\kappa_t = -1$) at $\sqrt{s}=13$~TeV (LO+MLM). The peak at $\Delta R \approx 3$--$4$ reflects the recoil of the forward jet against the central $tH$ system.}
\label{fig:5}
\end{figure}

The histogram in Figure~\ref{fig:5} shows the distribution of $\Delta R(p_j, h)$ (the angular separation between the forward light-quark jet and the Higgs boson) for simulated events in the $tHq$ final state, produced using the MadAnalysis~5 tool.

The $\Delta R$ spectrum exhibits a clear peak at $\Delta R \approx 3$--$4$, with a smooth rise from low values ($\Delta R = 0$--$2$) and a smooth fall-off at larger values ($\Delta R = 6$--$8$). This distribution is characteristic of the $tHq$ topology in the $t$-channel process: the Higgs boson is produced centrally (near $\eta \approx 0$), while the forward spectator jet recoils at high pseudorapidity ($|\eta_j| \sim 2$--$4$), resulting in a typical angular separation of $\sim 3$--$4$ units in $\Delta R$ space. This peaked structure is in excellent agreement with the expected kinematics of $t$-channel single-top production associated with a Higgs boson. In comparison to other channels analyzed in the ATLAS paper~\cite{atlas_tH_2025}:

\begin{itemize}
  \item In the 1-lepton channel ($1\ell$) with $H \to b\bar{b}$ and the 3-lepton channel ($3\ell$) with $H \to WW^*/\tau^+\tau^-/ZZ^*$, $\Delta R$ between key objects (including jet-Higgs pairs) contributes to BDT discrimination.
  \item The $tH$ signal is characterized by moderate $\Delta R$ values (peaked around 3--4), reflecting the recoil of the forward jet against the central $tH$ system.
  \item In contrast, $t\bar{t}+$jets backgrounds tend to peak at larger $\Delta R > 5$ due to multiple central jets and higher multiplicity, while $t\bar{t}W^\pm$ backgrounds often show softer peaks at lower $\Delta R \sim 2$--$3$ from more isotropic radiation.
\end{itemize}

The $\Delta R(p_j, h)$ distribution therefore provides strong discriminatory power against these backgrounds and is one of the important input features to the multivariate classifiers used in the ATLAS analysis. The observed peak at $\Delta R \approx 3$--$4$ in our LO+MLM simulations under the inverted Yukawa coupling hypothesis ($\kappa_t = -1$) is qualitatively consistent with the expected $t$-channel topology and with the post-fit distributions reported by ATLAS (e.g., in Section~9 and related figures for angular variables).

\subsection{$tWH$ process}

\begin{figure}[htbp]
\centering
\includegraphics[width=0.65\textwidth]{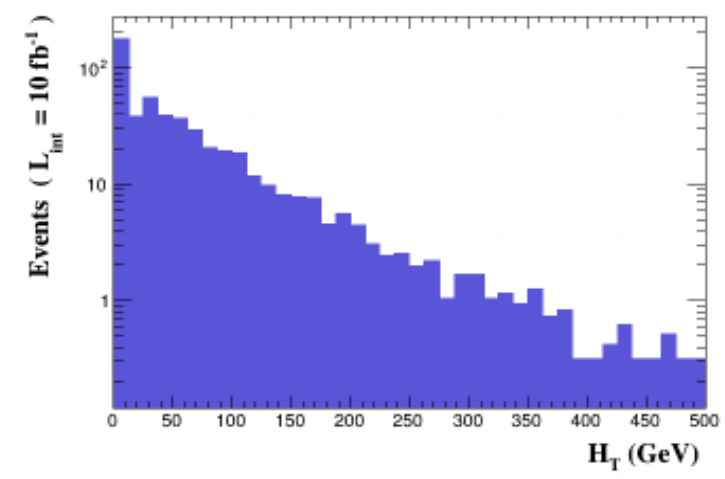} 
\caption{Distribution of the scalar sum of transverse momenta $H_T$ (in GeV) for simulated $tWH$ events in the inverted Yukawa coupling scenario ($\kappa_t = -1$) at $\sqrt{s}=13$~TeV (LO+MLM).}
\label{fig:6}
\end{figure}

Figure~\ref{fig:6} presents the distribution of the scalar sum of transverse momenta $H_T$ (in GeV) for simulated $tWH$ events in the inverted Yukawa coupling scenario ($\kappa_t = -1$), obtained from LO+MLM simulations.

The $H_T$ distribution shows a clear peak at $\sim$100 events per bin around $H_T \approx 100$--$150$~GeV, followed by a steep fall-off at higher values. This shape is typical of the $tWH$ process kinematics: the scalar sum is dominated by the transverse momenta of the top quark decay products, the $W$ boson (lepton + neutrino), the Higgs boson, and soft additional jets from initial- and final-state radiation. Phase-space constraints still limit very high $H_T$ values, but the extra electroweak gauge boson allows a broader tail than in the pure $t$-channel mode. The peak at moderate $H_T$ ($\sim$100--150~GeV) reflects the characteristic kinematics of associated $tWH$ production: the top and $W$ bosons are produced with moderate $p_T$, the Higgs is radiated with limited transverse momentum, and soft jets contribute modestly to the sum. The fall-off beyond $\sim$300--400~GeV is driven by PDF suppression at large momentum fractions and limited phase space for hard radiation.

This $H_T$ behavior is consistent with expectations for the $tWH$ mode and aligns qualitatively with the post-fit distributions in the ATLAS multi-lepton channels (e.g., $2\ell$SS and $3\ell$ targeting $H \to WW^*/\tau^+\tau^-$). As noted in Appendix~A of Ref.~\cite{atlas_tH_2025}, $H_T$ is a key BDT input variable for discriminating $tH$ signal from backgrounds such as $t\bar{t}W^\pm$ (which tends to have similar but slightly softer $H_T$) and $t\bar{t}+$jets (broader spectrum due to higher multiplicity).

\begin{figure}[htbp]
\centering
\includegraphics[width=0.55\textwidth]{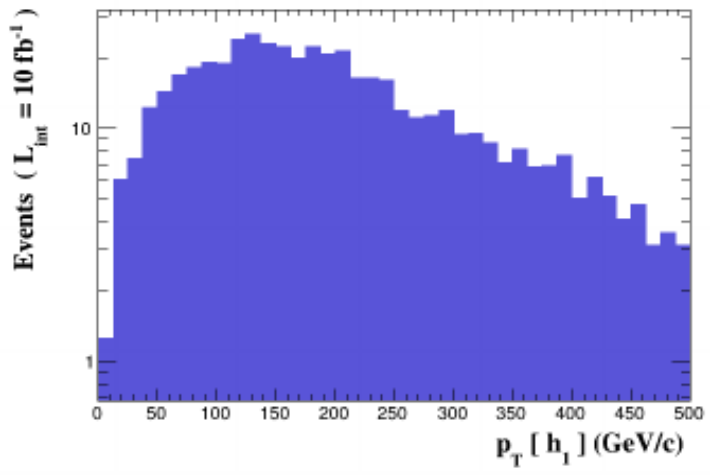} 
\caption{Transverse momentum distribution of the Higgs boson, $p_T(h)$, in GeV/c for simulated $tWH$ events in the inverted Yukawa coupling scenario ($\kappa_t = -1$) at $\sqrt{s}=13$~TeV (LO+MLM). The moderate peak and slightly broader low-$p_T$ shoulder reflect recoil against the top and $W$ bosons, with contributions from leptonic $W$ decay.}
\label{fig:7}
\end{figure}

The shape in Figure~\ref{fig:7} is typical for $tWH$ kinematics, where the Higgs transverse momentum $p_T(h)$ is moderate, arising primarily from recoil against the top quark and the $W$ boson. The distribution shows a broad peak at low-to-moderate $p_T(h)$ (typically 50--150~GeV/c) and tails that are suppressed by phase-space limitations at high values.

Compared to the $tHq$ channel, the $p_T(h)$ spectrum in $tWH$ is slightly broader at low $p_T$ due to additional kinematic contributions from the $W$ boson decay products (isolated lepton and neutrino). The leptonic $W \to \ell\nu$ decay adds transverse activity (via $E_T^{miss}$ and the lepton $p_T$), which effectively softens the low-$p_T(h)$ shoulder and extends the tail modestly compared to the cleaner $t$-channel recoil in $tHq$. This behavior is consistent with the expected topology of associated $tWH$ production: the top and $W$ bosons are produced with moderate transverse momenta, the Higgs is radiated with limited $p_T$, and the extra electroweak gauge boson allows greater flexibility in the overall event kinematics than in the pure $t$-channel mode. The suppression at very high $p_T(h)$ (beyond $\sim$300--400~GeV/c) reflects the same phase-space and PDF constraints as in $tHq$, but the presence of the $W$ decay products reduces the sharpness of the fall-off. The $p_T(h)$ distribution in the $tWH$ channel is an important input to the multivariate discriminants (BDTs) used in ATLAS multi-lepton analyses (e.g., $2\ell$SS and $3\ell$ channels targeting $H \to WW^*/\tau^+\tau^-$). It helps separate the $tH$ signal from backgrounds such as $t\bar{t}W^\pm$ (which tends to have similar but slightly harder $p_T(h)$ due to extra radiation) and $t\bar{t}+$jets (broader and more isotropic spectra). The moderate peak and extended low-$p_T$ shoulder in the $tWH$ mode provide complementary discrimination power to the forward-jet signatures in $tHq$.

\begin{figure}[htbp]
\centering
\includegraphics[width=0.55\textwidth]{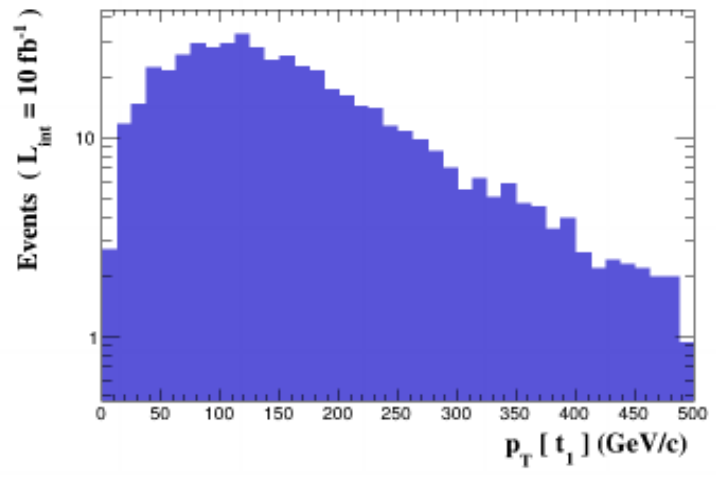} 
\caption{Transverse momentum distribution of the top quark, $p_T(t)$, in GeV/c for simulated $tWH$ events in the inverted Yukawa coupling scenario ($\kappa_t = -1$) at $\sqrt{s}=13$~TeV (LO+MLM). The peak at low-to-moderate $p_T(t)$ and steep fall-off at high values are typical of associated $tWH$ production.}
\label{fig:8}
\end{figure}

The transverse momentum distribution of the top quark, $p_T(t)$, in the $tWH$ process (shown in Figure~\ref{fig:8}) appears reasonable and consistent with typical kinematic expectations for this production mode at the LHC. Such distributions are well documented in the literature, including detailed simulations using tools like MadGraph5\_AMC@NLO at next-to-leading-order plus parton-shower (NLO+PS) accuracy.

In the Standard Model ($\kappa_t = +1$), the $p_T(t)$ spectrum generally peaks at low-to-moderate values (around 0--100~GeV/c), reflecting the relatively soft production of the top quark in associated $tWH$ events. The distribution falls off steeply toward higher $p_T$ (up to several hundred GeV/c), influenced by several factors:

\begin{itemize}
  \item the limited phase space available for hard top production in the presence of the additional $W$ boson and Higgs,
  \item destructive interference between diagrams involving Higgs-top Yukawa coupling and $W$-boson exchange (which suppresses high-$p_T$ tails),
  \item PDF suppression at large momentum fractions, and
  \item the recoil against the $W$ boson and Higgs radiation.
\end{itemize}

In the inverted Yukawa coupling scenario ($\kappa_t = -1$), the destructive interference turns constructive, enhancing the overall rate and mildly hardening the $p_T(t)$ tail compared to the SM case. This results in a slightly broader distribution and increased event fractions at moderate-to-high $p_T(t)$, which can improve signal efficiency in experimental selections and multivariate discriminants. The shape observed in Figure~\ref{fig:8}—a peak at low-to-moderate $p_T(t)$ and steep fall-off at higher values—is qualitatively consistent with these expectations and with post-fit distributions in the ATLAS multi-lepton channels (e.g., $2\ell$SS and $3\ell$ targeting $H \to WW^*/\tau^+\tau^-$). The top quark $p_T$ serves as a useful BDT input variable for discriminating $tH$ signal from backgrounds such as $t\bar{t}W^\pm$ (which tends to have similar but slightly harder $p_T(t)$ due to extra radiation) and $t\bar{t}+$jets (broader and more isotropic spectra).

\begin{figure}[htbp]
\centering
\includegraphics[width=0.55\textwidth]{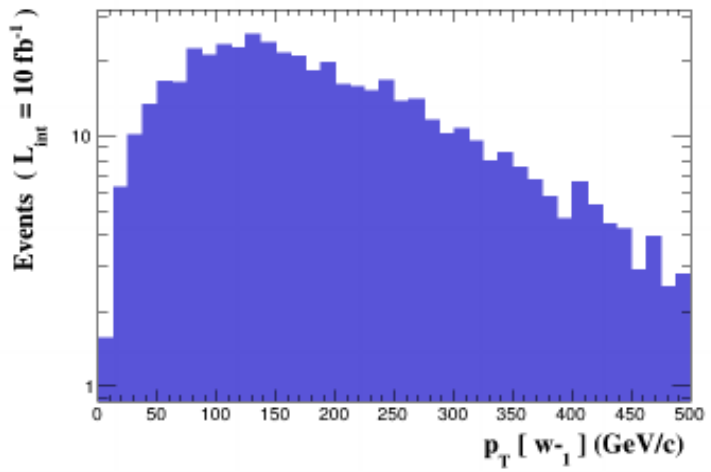} 
\caption{Transverse momentum distribution of the $W$ boson, $p_T(W)$, in GeV/c for simulated $tWH$ events in the inverted Yukawa coupling scenario ($\kappa_t = -1$) at $\sqrt{s}=13$~TeV (LO+MLM). The peak at low-to-moderate $p_T(W)$ and steep fall-off are typical of associated $tWH$ production.}
\label{fig:9}
\end{figure}

The transverse momentum distribution of the $W$ boson, $p_T(W^-)$ (or $p_T(W)$ collectively), in the $tWH$ process (shown in Figure~\ref{fig:9}) appears adequate and aligns well with kinematic expectations from theoretical studies at the LHC. Such distributions are well described in the literature, including detailed NLO+PS simulations presented in the 2017 paper on $tWH$ associated production~\cite{demartin2017}.

In the Standard Model ($\kappa_t = +1$), the $p_T(W)$ spectrum typically peaks at low values (around 0--100~GeV/c), reflecting the relatively soft production of the $W$ boson in associated $tWH$ events. The distribution exhibits a steep exponential fall-off toward higher $p_T$ (beyond 200--300~GeV/c), influenced by:

\begin{itemize}
  \item the limited phase space available for hard $W$ production in the presence of the top quark and Higgs boson,
  \item destructive interference between diagrams involving Higgs-top Yukawa coupling and pure electroweak $W$ exchange (which suppresses the high-$p_T$ tail),
  \item PDF suppression at large momentum fractions, and
  \item recoil effects against the top quark and Higgs radiation.
\end{itemize}

In the inverted Yukawa coupling scenario ($\kappa_t = -1$), constructive interference enhances the overall rate and mildly hardens the $p_T(W)$ tail compared to the SM case. This results in a slightly broader distribution and increased event fractions at moderate-to-high $p_T(W)$, which can improve signal efficiency in experimental selections, particularly in multi-lepton channels where the leptonic $W$ decay provides a clean signature (isolated lepton + $E_T^{miss}$). The shape observed in Figure~\ref{fig:9}---a peak at low-to-moderate $p_T(W)$ and steep fall-off at higher values—is qualitatively consistent with these theoretical expectations and with the kinematic distributions reported in Ref.~\cite{demartin2017} and related ATLAS post-fit plots in multi-lepton channels (e.g., $2\ell$SS and $3\ell$ targeting $H \to WW^*/\tau^+\tau^-$). The $W$ boson $p_T$ serves as a useful BDT input variable for discriminating $tH$ signal from backgrounds such as $t\bar{t}W^\pm$ (which tends to have similar but slightly harder $p_T(W)$ due to extra radiation) and diboson processes.

\section{Conclusions}

In the present work, we performed a thorough phenomenological analysis of the Higgs boson production in association with a single top quark at the LHC, with a particular focus on the sensitivity of the considered process to the size and sign of the top-Higgs Yukawa coupling. The $tH$ production modes, namely $tHq$ ($t$-channel) and $tWH$ (associated) processes, allow for a very powerful test of the relative phase between the Higgs couplings to fermions and gauge bosons due to the characteristic structure of the contributing amplitudes.

The interference pattern between diagrams involving the Higgs-top Yukawa coupling and those mediated by $W$-boson exchange makes the $tH$ cross-section and kinematic distributions highly sensitive to the sign of $\kappa_t \equiv y_t / y_t^{\rm SM}$. In the Standard Model ($\kappa_t = +1$), destructive interference suppresses the production rate, yielding $\sigma_{tH} \approx 89.5$~fb at NLO QCD for $\sqrt{s}=13$~TeV. In contrast, an inverted coupling ($\kappa_t = -1$) converts this into constructive interference, enhancing the cross-section by nearly an order of magnitude ($\sim$890~fb at NLO) and modifying the shapes of key observables (e.g., harder $p_T$ spectra, altered angular correlations).

This unique sensitivity, combined with the direct tree-level access to $y_t$ (unlike loop-induced processes such as ggF or $H\to\gamma\gamma$), positions $tH$ production as one of the most promising channels for probing non-standard Higgs-top interactions and potential CP-violating phases in the Yukawa sector. The following sections detail our simulation strategy, comparison with ATLAS Run-2 results, and the phenomenological implications of the inverted coupling hypothesis.

Utilizing state-of-the-art Monte Carlo simulations with MadGraph5\_aMC@NLO at LO+MLM accuracy, we have adopted a modeling strategy that is very close to the one used in the ATLAS Run-2 analysis. The four-flavour scheme adopted for $tHq$ and the five-flavour scheme adopted for $tWH$, along with dynamic scale choices and selection criteria motivated by the ATLAS results, enable a direct and controlled comparison with the experimental measurements.

Once appropriate $K$-factors are applied to approximate the NLO QCD corrections, the resulting prediction for the total $tH$ cross-section in the Standard Model at $\sqrt{s} = 13$~TeV is found to be in good agreement with the NLO theoretical expectation within the quoted scale and PDF uncertainties. This consistency validates the reliability of our LO+MLM simulation chain (including parton showering, PDF sets, and matching/merging procedures) as a practical tool for phenomenological studies of $tH$ production and its sensitivity to non-standard Higgs-top interactions. After normalization, our SM prediction yields $\sigma_{tH} \approx 85.9$~fb (with scale uncertainties $\sim\pm$10--15\% and PDF uncertainties $\sim\pm$4\%), which lies within the theoretical expectation of $\sim$89.5~fb reported in the ATLAS analysis and LHC Higgs Working Group recommendations.

A central outcome of this study is the explicit demonstration of the dramatic enhancement of the $tH$ production rate under an inverted top Yukawa coupling scenario ($\kappa_t = -1$). In this case, the destructive interference present in the Standard Model is converted into a constructive one, leading to an increase of the total cross section by several factors already at the LO+MLM level ($\sim$4--5 in our simulations), and by nearly an order of magnitude when extrapolated to NLO accuracy ($\sim$890~fb vs. 89.5~fb in SM). Our results qualitatively and quantitatively support the interpretation of the mild excess reported by ATLAS~\cite{atlas_tH_2025} in terms of scenarios with modified top–Higgs interactions.

In addition to the inclusive rates, we have performed a detailed study of key kinematic distributions: $H_T$, transverse momenta of the Higgs boson, top quark, and $W$ boson, pseudorapidity of the forward jet, and angular separations. These observables display the characteristic features expected for $tH$ production and exhibit clear sensitivity to the interference pattern dictated by $\kappa_t$. In particular, the forward jet signature in $tHq$ and the modified $p_T$ spectra in the inverted coupling scenario ($\kappa_t = -1$) enhance the discriminating power of the multivariate methods (BDTs) employed in current experimental analyses.

Overall, this work strengthens the theoretical foundation of $tH$ searches at the LHC by validating simplified LO+MLM calculations against NLO expectations and by elucidating the phenomenological consequences of non-standard Yukawa couplings. These results reaffirm the importance of $tH$ production as a precision probe of the Higgs sector and as a potential portal to physics beyond the Standard Model. Looking ahead, the substantially increased luminosity of the High-Luminosity LHC (HL-LHC), combined with improved theoretical control and advanced analysis techniques, is expected to enable a much more precise determination of the top–Higgs Yukawa coupling—including its possible CP-violating structure—thereby deepening our understanding of electroweak symmetry breaking.

\end{document}